\documentclass[eqsecnum,twocolumn,showpacs,amsfonts,aps,prc,floatfix]{revtex4} 

\usepackage{amsmath2000}
\usepackage{epsfig}

\newcommand\NPA{{Nucl. Phys.} A}

\newcommand\PLB{{Phys. Lett.} B}
\newcommand\PR{{Phys. Rep.}}

\newcommand\PRL{Phys. Rev. Lett.}
\newcommand\PRC{{Phys. Rev.} C}
\newcommand\PRD{{Phys. Rev.} D}

\newcommand\ZPC{{Z. Phys.} C}

\newfam\BMath
\font\BMathL=cmmib10 
\font\BMathl=cmmib7
\font\BMathm=cmmib5
\textfont\BMath=\BMathL \scriptfont\BMath=\BMathl
\scriptscriptfont\BMath=\BMathm

\newcommand\K{{\fam\BMath k}}
\newcommand\X{{\fam\BMath x}}

\newcommand\V{{\fam\BMath v}}
\newcommand\Z{{\fam\BMath z}}
\newcommand\bfat{{\fam\BMath b}}

\renewcommand\a{\alpha}

\newcommand\g{\gamma}
\renewcommand\d{\delta}

\newcommand\z{\zeta}
\newcommand\h{\eta}

\newcommand\m{\mu}
\newcommand\n{\nu}

\newcommand\p{\pi}
\newcommand\s{\sigma}
\renewcommand\t{\tau}

\newcommand\f{\phi}

\newcommand\cc{{\cal C}}

\newcommand\EQ{{\,=\,}}  

\newcommand\intks{\int \frac{\dd^3 \K}{(2\pi)^3}}

\newcommand{\lan}{\langle}     
\newcommand{\ran}{\rangle}     

\newcommand\del{\partial}

\newcommand{\nonum}{\nonumber}

\newcommand{\half}{\frac{1}{2}}

\newcommand\be{\begin{equation}}
\newcommand\ee{\end{equation}}
\newcommand\bea{\begin{eqnarray}}
\newcommand\eea{\end{eqnarray}}
\newcommand\beal{\begin{align}}
\newcommand\eeal{\end{align}}

\newcommand\ba{\begin{array}}
\newcommand\ea{\end{array}}
\newcommand\bc{\begin{center}}
\newcommand\ec{\end{center}}

\newcommand\eref[1]{Eq.~(\ref{#1})}
\newcommand\etwref[2]{Eqs.~(\ref{#1}) and (\ref{#2})}

\newcommand\bpi[1]{\begin{picture}#1}
\newcommand\epi{\end{picture}}

\newcommand{\fref}[1]{Fig.~\ref{#1}}

\newcommand{\tref}[1]{Table~\ref{#1}}

\def\jou#1#2#3#4{{#1} {\bf #2}, #3 (#4)}

\newcommand\B{{\fam\BMath b}}

\newcommand\ve{\varepsilon} 

\renewcommand\intks{\int \frac{d^3 k}{(2\pi)^3}}

\newcommand\kt{k_\perp}
\newcommand\ktv{\K_\perp}
\newcommand\hktv{\hat \K_\perp}

\newcommand\xt{x_\perp} 
\newcommand\xtv{\X_\perp}
\newcommand\vt{v_\perp} 
\newcommand\vtv{\V_\perp}

\newcommand\hvtv{\hat \V_\perp}

\newcommand\hzv{\hat \Z}
\newcommand\pt{P_\perp}
\newcommand\gt{\g_\perp}
\newcommand\mt{m_\perp} 

\newcommand\fo{{\text{f}}}

\begin{document}


\title{Elliptic Flow from a Transversally Thermalized Fireball}

\author{U. Heinz and S.M.H. Wong \\ }

\affiliation{Physics Department, The Ohio State University, Columbus,
 OH 43210 \\
}

\begin{abstract} 
The agreement of elliptic flow data at RHIC at central rapidity with 
the hydrodynamic model has led to the conclusion of very rapid thermalization. 
This conclusion is based on the intuitive argument that hydrodynamics,
which assumes instantaneous local thermalization, produces the largest
possible elliptic flow values and that the data seem to saturate this limit. 
We here investigate the question whether incompletely thermalized viscous
systems may actually produce more elliptic flow than ideal hydrodynamics. 
Motivated by the extremely fast primordial longitudinal expansion of the 
reaction zone, we investigate a toy model which exhibits thermalization 
only in the transverse directions but undergoes collisionless 
free-streaming expansion in the longitudinal direction. For collisions
at RHIC energies, elliptic flow results from the model are compared
with those from hydrodynamics. With the final particle yield and
$\kt$-distribution fixed, the transversally thermalized model is shown not 
to be able to produce the measured amount of elliptic flow.
This investigation provides further support for very rapid local kinetic 
equilibration at RHIC. It also yields interesting novel results for
the elliptic flow of massless particles such as direct photons. 
\end{abstract}

\vspace{1.0cm}

\date{\today}  

\maketitle

\section{Introduction} 
\label{s:intro}  

Recently the STAR collaboration showed \cite{st1,st2,rs} that the $\kt$ 
and centrality dependence of the measured elliptic flow coefficient $v_2$ 
of pions and protons at the Relativistic Heavy Ion Collider (RHIC) largely 
agreed with those generated from the hydrodynamic model simulation of 
heavy ion collisions \cite{o92,ksh,khh,hkhrv}. This was subsequently 
confirmed by data from the PHENIX \cite{lacey} and PHOBOS \cite{park} 
collaborations. $v_2$ is the second coefficient of an azimuthal Fourier
expansion of the transverse momentum spectrum around the beam
axis \cite{vz96} and, for collisions between identical nuclei, is the
lowest non-zero anisotropic flow coefficient at midrapidity. Given that 
the main prerequisite of the hydrodynamic model is complete local thermal 
equilibrium, which requires very intense rescattering among the matter 
constituents, it was argued that hydrodynamics should give the largest 
possible elliptic flow, and the observation that the data saturate the 
hydrodynamic limit was taken as evidence that thermalization must be very 
fast at RHIC for central and semi-central collisions \cite{khh,hkhrv,hk}. 

Elliptic flow {\em requires} reinteractions within the produced matter 
as a mechanism for transferring the initial spatial deformation of the 
reaction zone in non-central collisions onto momentum space. It it thus 
plausible to expect that the largest elliptic flow signal is produced in 
the hydrodynamic limit, i.e. in the limit of infinite rescattering rates
\cite{khh,hkhrv,hk}; however, a proof for this hypothesis has not yet been 
found. In hydrodynamics, collective flow is generated by pressure gradients, 
and flow anisotropies such as elliptic flow require anisotropic pressure 
gradients in the plane transverse to the beam direction. The longitudinal 
pressure plays a less visible role: the work done by the longitudinal 
pressure reduces the transverse energy $dE_T/dy$ and thereby, for given 
initial conditions for the energy density deposited in the reaction zone, 
the amount of transverse flow at freeze-out. If there were no longitudinal 
pressure, more of the initially deposited energy would go into the 
transverse directions, leading to flatter transverse momentum spectra. It 
is, at least in principle, conceivable that this could also lead, at fixed 
radial flow, to larger anisotropies of the transverse collective flow. If 
this were indeed the case and, at fixed slope of the angle-averaged spectra, 
larger values of $v_2$ could be generated in this way than within the usual 
hydrodynamic, the early thermalization argument would break down since 
the RHIC data \cite{st1,st2,rs} would no longer saturate the theoretical 
prediction from such a model. 

To explore this hypothetical possibility, we study in this paper a toy 
model which assumes that, at high collision energies, during the earliest 
collision stages the longitudinal momenta of the produced particles do 
not thermalize, but that the strong initial longitudinal expansion is 
instead dominated by collisionless free-streaming. The initial transverse
momenta are much smaller and assumed to thermalize quickly. This results  
in a system with local transverse, but zero longitudinal pressure. Due
to the masslessness of the partons created in the reaction zone the 
trace of the energy momentum tensor vanishes; in the absence of 
longitudinal pressure $P_\parallel$, the transverse pressure $P_\perp$
must thus be related to the initial energy density $\ve\EQ{T}^{00}$ by
$\ve\EQ2P_\perp$ (instead of the usual $\ve\EQ3P$). This results
in a much stiffer equation of state for the transverse dynamics (see
Sec.~\ref{s:tmn}), giving more flow and possibly stronger flow 
anisotropies. Of course, if the transverse momenta thermalize, the 
longitudinal momenta should do so eventually, too. We assume in our 
model that when this happens the flow anisotropies have already 
almost reached their asymptotic values \cite{fn1}.

Similar to the hydrodynamic simulations \cite{o92,ksh,khh,hkhrv}, we 
assume that the longitudinal free-streaming dynamics is boost invariant 
\cite{bj} and correspondingly concentrate on the central rapidity 
region. Starting from a kinetic description of the longitudinally 
free-streaming but transversally thermalized gluonic system (which we 
call ``transversally thermalized model'', TTHM), we derive a set of 
macroscopic evolution equations (TTHM equations) which we solve with 
appropriate initial conditions. The impact parameter dependence of 
the initial conditions is handled in the same way as in the hydrodynamic
simulations. When comparing our TTHM solutions to those from the ideal 
hydrodynamic model (HDM) we retune the initial conditions in such a way 
that for central collisions ($b=0$\,fm) roughly the same multiplicities 
and spectral slopes at midrapidity are obtained. We then compare the 
momentum-space anisotropy $v_2$ in the two approaches, as a function of 
transverse momentum $k_\perp$ and impact parameter $b$. 

\section{Constructing a Transversally Thermalized System} 
\label{sec2}
\subsection{The Phase-Space Distribution Function}
\label{sec2a}

We consider a gluon dominated system at a short time after nuclear contact.
We assume that at that time the transverse gluon momenta are already 
thermalized but that the system is still free-streaming along the beam 
direction. The system is required to possess longitudinal boost invariance, 
reflecting a boost invariant primary particle production mechanism \cite{bj}. 
For easiest implementation we assume that the collision energy is so high
that the colliding nuclei are Lorentz contracted to two infinitesimally 
thin sheets in the $z$-direction and that all produced particles point 
back to $z=t=0$. Their longitudinal momenta and coordinates thus satisfy 
$z=u_z t$ or $k_0 z =k_z t$ where $u_z=k_z/k_0$. This automatically 
identifies \cite{bj} the rapidity 
 \be  
   y = \half \ln \Big (\frac{k_0+k_z}{k_0-k_z} \Big )
 \ee
of the produced particles with their spacetime rapidity
 \be  
   \h = \half \ln \Big (\frac{t+z}{t-z} \Big ) .
 \ee
We thus make the ansatz 
 \be
   f(\K,\X,t) = \d (k_0 z-k_z t) \; \tilde h(\K,\X,t) \;. 
 \label{eq:an}
 \ee 
Longitudinal free-streaming is implemented by the condition
 \be 
   \left (k_0 \frac{\del}{\del t} +k_z \frac{\del}{\del z} \right ) 
     f(\K,\X,t)  = 0  \;. 
 \ee 
Since the $\d$-function factor solves this equation trivially, this implies
a constraint for $\tilde h$. The latter is most conveniently expressed by
using an alternate coordinate system (we hereafter make explicit use 
of the masslessness of the produced gluon degrees of freedom, by setting
$\mt=\sqrt{m^2+\kt^2}\to \kt$):  
 \begin{eqnarray}
     t &=& \t   \cosh \h             \;, \qquad \;\;\; 
     z = \t   \sinh \h             \;,                  \\
     k_0 &=& \kt \cosh y   \;, \qquad
     k_z = \kt \sinh y   \;.
 \end{eqnarray}
In these coordinates the ansatz \eref{eq:an} becomes 
\be
   f(\K,\X,t) = \frac{\d (\h-y)}{\kt \t \cosh (\h-y)} 
                 \; \tilde h(\ktv,\xtv,\h-y,\t) 
 \label{fnew}
\ee
where the dependence of $\tilde h$ on only $\ktv$ and the difference 
$\h-y$ results from the requirement of longitudinal boost invariance. 
We implement transverse thermalization by the ansatz
 \be
    \tilde h(\ktv,\xtv,\h{-}y,\t) = 
    \frac{\t_0 \kt}{\gt} \cosh (\h{-}y)\; \frac{1}{e^{k\cdot u/T}-1},
 \label{eq:tilh}
 \ee
where $u^\m{\,=\,}\gt (\cosh \h, \vtv, \sinh \h)$ is the flow vector with
Bjorken longitudinal flow velocity $v_z{\,=\,}\tanh \h$ \cite{bj} and 
transverse flow velocity $\vtv{\,=\,}\vtv(\xtv,\t)$ 
($\gt^{-1}{\,=\,}\sqrt{1{-}\vt^2}$).
$T=T(\xtv,\t)$ is the temperature characterizing the thermalized 
transverse momentum spectrum. In the local rest frame $T$ is a function 
of the rest frame coordinates $\xtv^*$ and $\t^*{\,=\,}t^*$ which after the 
standard \cite{ruu} successive transverse and longitudinal boosts to the
laboratory frame, with $\vtv$ and $v_z{\,=\,}\tanh \h$ respectively, are 
related to the laboratory variables by 
 \bea
    t^* &=& \gt (t \cosh \h -z \sinh \h -\vtv \cdot \xtv )  \nonum\\   
        &=& \gt (\t -\vtv \cdot \xtv ) = u\cdot x                
 \label{t*}                                                       \\
    {\xt^*}_{\parallel} &=& \gt ({\xt}_{\parallel} -\vt (t\cosh\h -z\sinh\h))
                                                            \nonum\\
        &=& \gt ({\xt}_{\parallel} -\vt \t) = -m\cdot x
 \label{x*par}                                                    \\
  {\xt^*}_{\perp} &=& {\xt^{}}_{\perp} \;.                        
 \label{x*perp} 
 \eea
The additional $\parallel$ and $\perp$ subscripts denote the components 
of $\xtv$ parallel and perpendicular to $\vtv$, respectively, and
the vector $m^\m$ is defined in \eref{eq:vec} below. The sequence of 
boosts described here agrees with the standard choice \cite{ruu} and 
gives the simplest structure for the set of 4-vectors needed for 
decomposition of the energy momentum tensor (see Sec.~\ref{s:tmn}).
As one can see from Eqs.~(\ref{t*})-(\ref{x*perp}), the dependence of
$T$ on $\xtv^*$ and $t^*$ in the local rest frame translates into 
a dependence on $\xtv$ and $\t$ in the laboratory frame.

By virtue of the factor $\d(\h{-}y)$, the argument of the Bose distribution 
in \eref{eq:tilh} reduces in the local rest frame to $k{\cdot}u\EQ\kt$, 
showing explicitly that the system is only transversally thermalized. 
Combining \etwref{fnew}{eq:tilh} and exploiting the $\d$-function, the 
distribution function simplifies to 
\bea  
\label{eq:dist}
  &&f(\K,\X,t) = \frac{\t_0}{\t} \;\d(\h-y)\,h(\ktv,\xtv,t)\,,
\nonum\\
  &&h(\ktv,\xtv,t) = \frac{1}{\gt}\, \frac{1}{e^{k\cdot u/T}-1}  \;. 
\eea
In the new coordinates, $f$ satisfies the relativistic transport equation 
 \bea  
   &&\Bigl( \kt \cosh (\h{-}y) \frac{\del}{\del \t} 
            -\frac{\kt}{\t} \sinh (\h{-}y) \frac{\del}{\del \h} 
 \nonumber\\
   &&\qquad\quad   +\,\ktv{\cdot}\nabla_\perp 
    \Bigr) f(\K,\X,t) = \cc(\K,\X,t)     \;, 
 \label{eq:tr-eq} 
 \eea
where the collision term $\cc$ is responsible for keeping the transverse
momenta thermalized while not changing any of the longitudinal momenta. 
Due to the implemented free-streaming properties we have
\be  
    \left ( \cosh (\h{-}y) \frac{\del}{\del \t} 
            -\sinh (\h{-}y) \frac{1}{\t} \frac{\del}{\del \h} 
    \right ) \left(\frac{\t}{\t_0}\d(\h-y)\right) = 0 \nonum
\ee
such that \eref{eq:tr-eq} reduces to the following kinetic equation for 
the transverse distribution function:
\be  
    \left ( \kt \frac{\del}{\del \t} 
            +\ktv \cdot \nabla_\perp 
    \right ) h(\ktv,\xtv,\t) = \cc(\K,\X,t). 
\label{eq:tr-tr} 
\ee
The space-time evolution of $h(\ktv,\xtv,\t)$ is entirely due to the
collisions and collective transverse expansion, but decoupled
from the boost invariant longitudinal expansion. The collisions must 
be sufficiently dominant over the transverse expansion to maintain the 
equilibrium form of $h(\ktv,\xtv,\t)$ given above. 

We are aiming at a macroscopic description on the basis of the 
differential conservation laws for energy and momentum, similar to
hydrodynamics. This requires the construction of $T^{\m\n}$ from
the distribution functions. Due to the different microscopic physics
in the longitudinal and transverse directions implemented in our model,
there are now more vectors required to construct a complete set of
tensors with respect to which $T^{\m\n}$ should be decomposed. 
The set of four-vector fields that we need are 
\bea u^\m (\xtv,\h,\t)  & = & \gt (\cosh \h, \vtv, \sinh \h),  \nonum \\ 
     n^\m (\xtv,\h,\t)  & = &     (\sinh \h, 0,0, \cosh \h), \\ 
     m^\m (\xtv,\h,\t)  & = & \gt (\vt \cosh \h, \hvtv, \vt \sinh \h).\nonum 
\label{eq:vec} 
\eea
They arise from the following local rest frame vectors by successive
boosts with $\vtv$ and $v_z$ as described above:
\bea u^\m (\xtv,\h,\t)  & = & (1,0,0,0)\;,  \nonum \\ 
     n^\m (\xtv,\h,\t)  & = & (0,0,0,1)\;,  \nonum \\ 
     m^\m (\xtv,\h,\t)  & = & (0,\hvtv,0) \;. 
\label{eq:vec_r} 
\eea
One is timelike, $u^2{\EQ}1$, while the other two are spacelike,
$n^2{\EQ}m^2{\EQ}{-}1$. They are mutually orthogonal: 
$u{\cdot}n{\EQ}$ $u{\cdot}m{\EQ}n{\cdot}m{\EQ}0$. With these vectors, 
the distribution can be expressed in a form which is explicitly 
boost invariant under longitudinal boosts:
\be  
     f(\K,\X,t) = \frac{\t_0}{\t} \; 
                  \frac{\d(k\cdot n)\, (k\cdot u-\vt k\cdot m)} 
                       {e^{k\cdot u/T}-1}\;.  
\label{eq:dist-cov}
\ee 

\subsection{Energy-Momentum Tensor and Conservation Laws} 
\label{s:tmn}

Based on the structure (\ref{eq:dist-cov}) for the phase-space 
distribution function we want to derive a macroscopic set of dynamical
equations based on the conservation laws for energy and momentum. 
In terms of the distribution function $f$, the energy-momentum tensor
is given by 
\be 
   T^{\m\n} (\X,t) = \n_g \intks \frac{k^\m k^\n}{k^0} f(\K,\X,t) \;,
\label{eq:tmn-def}
\ee 
where $\n_g\EQ2\cdot8\EQ16$ is the degeneracy factor for color and helicity 
of the gluons. From the previous subsection we know that $f(\K,\X,t)$ 
depends on the three 4-vectors $u^\m$, $n^\m$ and $m^\m$. The most general 
form of the tensor $T^{\m\n}$ thus reads 
 \bea 
   &&T^{\m\n} = A u^\m u^\n +B n^\m n^\n +C m^\m m^\n
                 +D (u^\m n^\n{+}n^\m u^\n)                   
 \nonum \\
   &&\ +F (u^\m m^\n{+}m^\m u^\n) +G (n^\m m^\n{+}m^\m n^\n) 
                 +H g^{\m\n}. 
 \label{eq:tmn-ten}
 \eea
Since the gluons are massless, the trace of  $T^{\m\n}$ vanishes:
\be T^\m_\m = A -B -C +4H = 0 \;. 
\ee 
In the local rest frame we have 
\bea 
\label{rest}
     T^*_{00} &=& A+H    = \ve\;, \nonum\\
     T^*_{xx} &=& T^*_{yy} = -H = \pt \;, \\
     T^*_{zz} &=& B-H    = P_z\,, \nonum     
\eea
where $\ve$ is the energy density and $\pt$ and $P_z$ are the transverse 
and longitudinal pressure, respectively. In a general frame, all the 
Lorentz scalar coefficients in Eq.~(\ref{eq:tmn-ten}) can be found by 
contracting $T^{\m\n}$ with all pairwise combinations of the 4-vectors 
$u^\m$, $n^\m$ and $m^\m$. We will denote these contractions by 
$(uTn){\,\equiv\,}u_\m T^{\m\n} n_\n$ etc. The details of this calculation 
are given in Appendix~\ref{appa}) where we find
\begin{align}
\label{eps}
 (uTu) & = A+H \!\!\!\!& = &\,\frac{\t_0}{\t} \n_g \frac{\p^2 T^4}{60}
               \!\!\!\!& = & \; \ve \\
 (uTm) & =-F \!\!\!\!  & = &-\frac{\t_0}{\t} \n_g \frac{\p^2 T^4}{120} \vt 
       \!\!\!\!& = & -\frac{\vt}{2}\ve \\
 (mTm) & = C-H\!\!\!\! & = &\,\frac{\t_0}{\t} \n_g \frac{\p^2 T^4}{120} 
       & = & \; \half\ve \\    
 (uTn) & =-D   & = &\,0                                    &   & \;  \\ 
 (nTn) & = B-H\!\!\!\! & = &\,0                            & = & \; P_z \\
 (nTm) & = G   & = &\,0       
\label{nTm}           
\end{align}
From this and Eq.~(\ref{rest}) we see that 
\be 
    B = H = -\pt\;,\qquad A = \ve + \pt\;,
\ee
and the tracelessness condition gives
\be
    \ve + \pt +\pt -\half \ve -3\pt = \half \ve -\pt = 0 \;. 
\ee
Therefore the equation of state is 
 \be
 \label{eos} 
    \ve = 2 \pt \;. 
 \ee 
Combining this with the expression for $(mTm)$ one finds $C=0$. 
The expression for the energy momentum tensor is then
\bea
\label{tmn} 
   T^{\m\n} &=& (\ve+\pt) u^\m u^\n -\pt n^\m n^\n -\pt g^{\m\n} 
\nonum\\
            &&+\pt \vt (u^\m m^\n+m^\m u^\n)  \;. 
\eea 

For completeness and later use we also give the expression for the gluon
density in the local comoving frame:
\be
\label{gdens}
  n_g = u\cdot j = \frac{\t_0}{\t}\n_g \frac{\zeta(3) T^3}{2\pi^2},
\ee
where $j^\m(x)$ is the gluon number current:
\be 
\label{j-def}
  j^\m (\X,t) = \n_g \intks \frac{k^\m}{k^0} f(\K,\X,t) \;.
\ee  

We now proceed to derive the macroscopic evolution equations. We
first rewrite the conservation laws
\be 
  \del_\n T^{\m\n} = 0  
\ee
in $(\h,\t)$ coordinates,
\bea
\label{EOM} 
   \cosh\h \left(\frac{\del T^{0\m}}{\del\t}
    +\frac{1}{\t} \frac{\del T^{z\m}}{\del\h}\right) 
    &{-}&\sinh\h \left(\frac{\del T^{z\m}}{\del\t}
    +\frac{1}{\t} \frac{\del T^{0\m}}{\del\h}\right) 
 \nonum\\   
   &{+}&\nabla_\perp^i T^{i\m} = 0 \;,
\eea 
where the sum over $i$ goes over the two transverse directions. 
When writing these out explicitly in terms of $\ve$, $\pt$, and
the longitudinal and transverse flow velocities, using Eqs.~(\ref{tmn}) 
and (\ref{eq:vec}), one finds that all explicit dependence on $\h$
disappears. Boost-invariant initial conditions (i.e. $\h$-independent
initial expressions for $\ve$ and $P_\perp$ and the structure 
(\ref{eq:vec}) for the vectors $u^\m,\,n^\m,$ and $m^\m$) are thus 
preserved by the macroscopic evolution equations. 

For the further analysis we can thus concentrate on the dynamics in 
the central transverse plane at $z\EQ\h\EQ0$. There we have 
\bea u^\m &=& \gt (1,\vtv,0)\,,\quad 
     \frac{\del u^\m}{\del \h} = \gt (0,0,0,1)\,,             \\
     n^\m &=&     (0,0,0,1)\,,\qquad
     \frac{\del n^\m}{\del \h} =     (1,0,0,0) \,,            \\
     m^\m &=& \gt (\vt, \hvtv, 0)\,,\ 
     \frac{\del m^\m}{\del \h} = \gt (0,\hvtv,\vt)\,.    
\eea 
Using the $\h$-independence of $\ve$ and $\pt$ we find
 \bea 
    \frac{\del T^{z\m}}{\del\h} 
    &=& \g_\perp^2 (\ve+\pt) (\d^{0\m}+\vt^i \d^{i\m}) -\pt \d^{0\m} 
 \nonum\\
    &&+\g_\perp^2 \pt \Big (2\vt^2\d^{\m0} 
       +(1 + \vt^2)\vt^i\d^{i\m} \Big )              
 \nonum\\
    &=& T^{00} \d^{0\m} + T^{0i} \d^{i\m} = T^{0\mu}\,,                               
 \eea 
and the equations of motion (\ref{EOM}) at $z\EQ\h\EQ0$ reduce to 
\bea 
\label{t00}
      \frac{\del T^{00}}{\del \t} +\frac{T^{00}}{\t}  
    + \nabla_\perp^j T^{j0} & = & 0\,,
\\
\label{t0i}
      \frac{\del T^{0i}}{\del \t} + \frac{T^{0i}}{\t}  
    + \nabla_\perp^j T^{ji} & = & 0  \;. 
\eea 
Using the equation of state (\ref{eos}) to eliminate $\ve$, these three
equations completely determine the three unknown functions $\pt$ and $\vtv$.
Since gluons possess no conserved charges, no further current conservation 
laws need to be considered.

\section{Hydrodynamics}
\label{sec3}

For later comparison we here shortly review the parallel procedure
for ideal hydrodynamics based on a locally fully thermalized 
phase-space distribution function
\be
\label{equil}
  f(\K,\X,t) = \frac{1}{e^{k\cdot u/T}-1}\,.
\ee
In this case the energy-momentum tensor has the ideal fluid decomposition 
\be 
  T^{\m\n} = (\ve + P) u^\m u^\n -P g^{\m\n}\,,
\ee 
and for a longitudinally boost-invariant system $u^\m$ has the same form
as in \eref{eq:vec}. The equation of state is in this case 
\be
\label{eps_h}
 \ve = 3 P = \n_g \frac{\p^2 T^4}{30}\,, 
\ee
and the gluon density in the local rest frame is
\be
\label{gd}
  n_g = \n_g \frac{\zeta(3) T^3}{\pi^2}.
\ee
Note that the the energy per particle 
\be
\label{epp}
  \frac{\ve}{n_g} = \frac{\pi^4}{30\zeta(3)} T \approx 2.7\,T
\ee 
agrees exactly with the corresponding expression in the TTHM, see 
Eqs.~(\ref{eps}),(\ref{gdens}).

The pressure is now locally isotropic. With longitudinally boost-invariant
($\h$-independent) initial conditions for $\ve$ and $P$ the ansatz 
\eref{eq:vec} for the flow velocity is preserved in time by the 
equations of motion $\del_\n T^{\m\n}\EQ0$ \cite{bj}, and the latter
become
\bea  
\label{T00}
      \frac{\del T^{00}}{\del \t} +\frac{T^{00}}{\t}  
    + \nabla_\perp^j T^{j0} & = & -\frac{P}{\t}                
\label{eq:hdm-ce}
\\
\label{T0i}
      \frac{\del T^{0i}}{\del \t} + \frac{T^{0i}}{\t}  
    + \nabla_\perp^j T^{ji} & = & 0  \;. 
\eea  
The main difference between these hydrodynamical equations and our TTHM 
toy model equations (\ref{t00}) and (\ref{t0i}) is the pressure term on
the right hand side of Eq.~(\ref{T00}) which is absent in Eq.~(\ref{t00}).
It indicates the longitudinal work done by the isotropic pressure in 
hydrodynamics; in the transversely thermalized (TTHM) model there is no 
longitudinal pressure which could do work in the longitudinal direction.

\section{Initial Conditions}
\label{sec4}

We consider Au+Au collisions at RHIC energies. The initial energy
density distribution in the transverse plane as a function of impact 
parameter $\B$ is taken to be proportional to the density of binary 
nucleon collisions in the transverse plane, calculated from the 
normalized nuclear thickness function:
\be 
   T_A (\xtv) = \int \rho_A (\xtv,z)\, dz  ,\ \  
   \int d^2 \xt T_A(\xtv) = A . 
\ee 
For the nuclear number density we took the Wood-Saxon parametrization 
\be 
   \rho_A(\xtv) = \frac{\rho_0}{e^{(r-R_0)/\xi}+1}  
\ee
with $R_0\EQ1.14\,A^{1/3}$\,fm and $\xi\EQ0.54$ fm. For collisions 
between nuclei of masses $A$ and $B$ the number of binary nucleon 
collisions per unit area in the transverse plane is
\be 
   \frac{dN(\xtv,\B)}{d^2 \xt} = \s_0 T_A(\xtv) T_B(\B-\xtv) \;,
\ee
where $\s_0\EQ40$\,mb is the total inelastic nucleon-nucleon cross 
section. We assume that the generated energy density is proportional
to this density. We performed two classes of calculations: in the first 
class we used the same maximum initial energy density as the authors of 
\cite{khh,hkhrv} ($\ve_0\EQ{e}_0{\,\equiv\,}23.0$ GeV/fm$^3$ for central 
($b\EQ0$\,fm)) collisions at an initial time $\t_0\EQ0.6$ fm/$c$) in 
order to compare the HDM and TTHM results for similar initial conditions. 
In the second class of simulations we retuned the initial conditions in 
TTHM such that the same multiplicity density $dN/dy$ and the same slope 
for the transverse momentum spectrum at midrapidity as in HDM is obtained. 
Both sets of results will be discussed in the next Section.

\section{Elliptic Flow in a Transversally Thermalized System} 
\label{sec5}
\subsection{Freeze-out prescription}
\label{sec5a}

Just as the HDM, our TTHM model describes the time evolution of 
macroscopic thermodynamic quantities which must be converted to 
particle spectra before one can compare with experiments. To do so we 
use the well-known Cooper-Frye prescription \cite{cf} which gives
the particle spectrum in terms of an integral over the phase-space 
distribution function along a so-called freeze-out hypersurface $\Sigma(x)$: 
\be   
   E \frac{dN}{d^3k} = \frac{dN}{dy\,\kt d\kt d\f} 
   = \frac{\n_g}{(2\p)^3} \int_\Sigma k{\cdot}d^3\Sigma(x)\,f(x,k) \,.
\label{eq:cf} 
\ee 
$d^3 \Sigma_\m(x)$ is the four-vector integral measure normal to the 
hypersurface, and in the TTHM case $f(x,k)$ is a transversally thermalized 
distribution function of the form (\ref{eq:dist}), with flow velocity
$u^\mu$ evaluated along the freeze-out hypersurface $\Sigma$ and 
temperature $T$ calculated via Eq.~(\ref{eps}) from the energy density 
on $\Sigma$. Longitudinal boost-invariance dictates freeze-out along a 
surface of fixed longitudinal proper time $\t_\fo(\xtv)$, and we can 
write
\be
\label{sigma}
  \Sigma_\m(x) = \bigl(\t_\fo(\xtv)\cosh\h,\xtv,\t_\fo(\xtv)\sinh\h\bigr).
\ee
This gives
\be 
   k\cdot d^3 \Sigma = 
   \Big (\kt\cosh(y{-}\h) -\ktv\cdot\nabla_\perp \t_\fo\Big )
   \t_\fo\, d\h\, d^2 \xt  \;. 
\label{fo-norm}
\ee
Of course, the $\h$-integration is trivial due to the factor $\d(\h{-}y)$ 
in the distribution function (\ref{eq:dist}). 

\subsection{TTHM vs. HDM for identical initial conditions}
\label{sec5b}

\begin{figure}
\epsfig{file=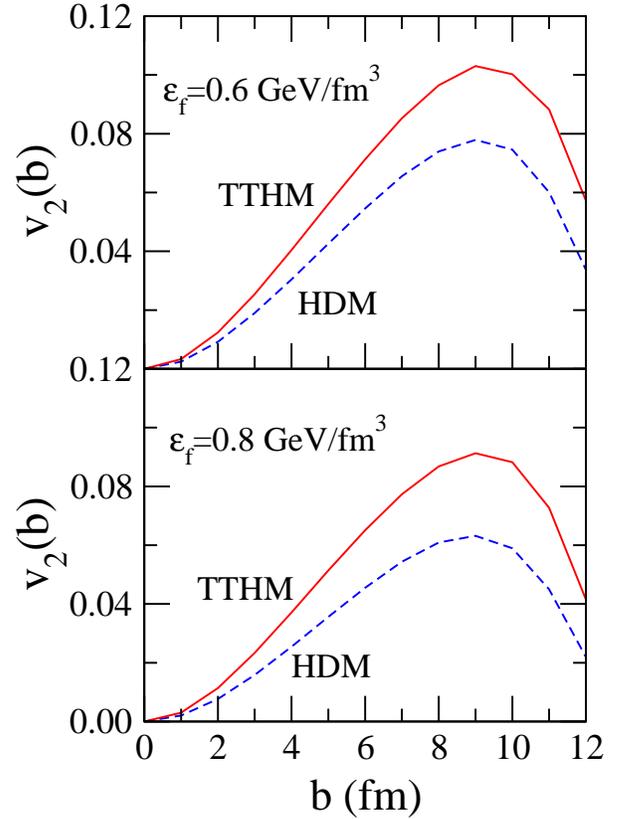,width=0.9\linewidth}
\caption{$v_2$ as a function of impact parameter at two freeze-out energy
densities $\ve_{\rm fo}$. The solid lines are from our TTHM model while 
the dashed lines are from hydrodynamic simulations with identical 
initial conditions.}  
\label{F1} 
\end{figure}

In \fref{F1} we compare results for the elliptic flow from the TTHM 
and HDM models for identical initial conditions, taken from \cite{khh,hkhrv}.
Since the temperature parameter $T$ has a different meaning in our model 
than usual (only the transverse momenta are thermalized), we enforce
``freeze-out'' (i.e. we stop the dynamical evolution and calculate the
spectra and momentum space anisotropies) along a surface of constant 
energy density $\ve_\fo$. We show results for two values of this 
parameter both of which lie in the quark-hadron transition region. 
Since hadronization certainly involves longitudinal momentum exchange, 
it will violate our assumption of longitudinal free-streaming, and we 
should definitely not follow the TTHM dynamics beyond the hadronization 
point. Of course, by truncating the dynamics at such high freeze-out 
energy densities we forfeit the possibility of comparing our results 
directly with experiment. However, a large fraction of the elliptic flow 
should already have developed at this point \cite{ksh,fn2}, and we can still 
make a meaningful comparison between the hydrodynamic evolution of a fully 
thermalized quark-gluon plasma and that of a longitudinally free-streaming 
gluonic system with only transverse thermalization, by comparing the
gluon spectra at this common ``final'' energy density $\ve_\fo$.

Figure \ref{F1} shows $v_2$ as a function of the impact parameter $b$. 
$v_2(b)\EQ\langle\cos(2\phi)\rangle(b)$ is computed from the $\kt$-integrated
gluon spectrum at freeze-out as
\bea v_2(b) &=& 
     \frac{\int \kt d\kt d\f \cos(2\f) \,\frac{dN}{dy\,\kt d\kt d\f}(b)} 
          {\int \kt d\kt d\f \frac{dN}{dy\,\kt d\kt d\f}(b)}
\nonum\\ 
      &=& \frac{\int d\f \cos(2\f) \frac{dN}{dy d\f}(b)} 
                  {\int d\f \frac{dN}{dy d\f}(b) }              \;. 
\eea
The Figure shows that, {\em for identical initial conditions}, the TTHM 
dynamics generates a {\em larger} momentum-space asymmetry $v_2$ than 
hydrodynamics. We also see that $v_2$ has not yet fully saturated at 
these values of $\ve_\fo$; the reason for this can be found in 
Table~\ref{T1} which shows that, with the very hard equation of 
state and initial conditions used here, the above values of $\ve_\fo$ 
are reached quite early, before the initial spatial deformation has been 
fully eliminated (see also Fig.~1 in \cite{ksrh}).

\begin{table}[t]
\caption{\label{T1} Rapidity densities at $y\EQ0$ and final times
$\t_\fo$ for central collisions with identical initial conditions (see text)
in the TTHM and HDM simulations.\\} 
\begin{ruledtabular}
\begin{tabular}{c|cc|cc} 
                          &  \multicolumn{2}{c}{TTHM}         \vline 
                          &  \multicolumn{2}{c}{HDM}         \\ \cline{2-5}
  \ $\ve_\fo$ (GeV/fm$^3$)\    
                          &\  $dN/dy$      & \hspace{0.1cm} $\t_\fo$ (fm/$c$) 
                                             \hspace{0.1cm}
                          &\  $dN/dy$      & \hspace{0.1cm} $\t_\fo$ (fm/$c$) 
                                             \hspace{0.1cm} \\ \hline 
  0.6                     &\  780      &  6.95  
                          &\  944      &  5.97              \\ 
  0.8                     &\  780      &  6.48    
                          &\  940      &  5.36              \\  
\end{tabular}
\end{ruledtabular}
\end{table} 

\begin{figure}
\epsfig{file=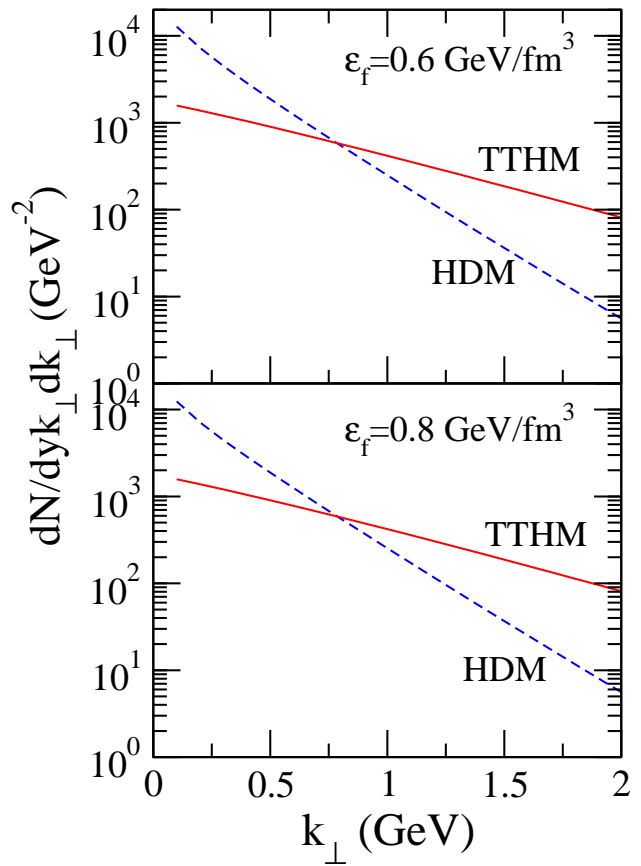,width=0.95\linewidth}
\caption{Gluon $\kt$ distribution for central collisions ($b=0$\,fm) and
two different values for the freeze-out energy density, for identical initial
conditions. The solid lines are for our model while the dashed lines 
are for hydrodynamics.}  
\label{F2} 
\end{figure}

The larger values of $v_2$ from TTHM, however, come also with much flatter
transverse momentum spectra, as shown in Fig.~\ref{F2}. Part of this 
flattening is due to a higher freeze-out temperature in the TTHM. By 
comparing Eqs.~(\ref{eps}) and (\ref{eps_h}) one sees at the same energy
density $\ve_\fo$ the temperatures in the two models are related by
$T_{\rm f,T}=(2\t_\fo/\t_0)^{1/4}\,T_{\rm f,H}$. With the $\t_\fo$ 
values given in Table~\ref{T1}, the TTHM freeze-out temperature is about 
a factor 2 higher than the HDM one. However, the inverse slopes of the 
spectra shown in Fig.~\ref{F2} differ by more than this factor 2 (the 
difference is closer to a factor 2.5-2.6). This implies that TTHM also 
creates somewhat stronger radial flow than HDM which, in view of its 
harder equation of state, is not unexpected.

The initial conditions for the hydrodynamic model (in particular the 
initial energy density) were tuned \cite{khh} to reproduce the measured 
final multiplicity densities $dN_{\rm ch}/dy$ and spectral slopes in 
central Au+Au collisions at RHIC; clearly, with these same initial 
conditions, the transversally thermalized model TTHM will no longer be 
anywhere close to these data. The larger $v_2$ values from TTHM in 
Fig.~\ref{F1} therefore cannot be compared with the RHIC 
data, and the comparison with HDM in Fig.~\ref{F1} is misleading. For a 
meaningful comparison the initial conditions for TTHM should first be 
adjusted in such a way that at least the multiplicity densities and 
spectral slopes for central collisions are similar in both models. This
will be our next step.

\begin{figure}[t]
\epsfig{file=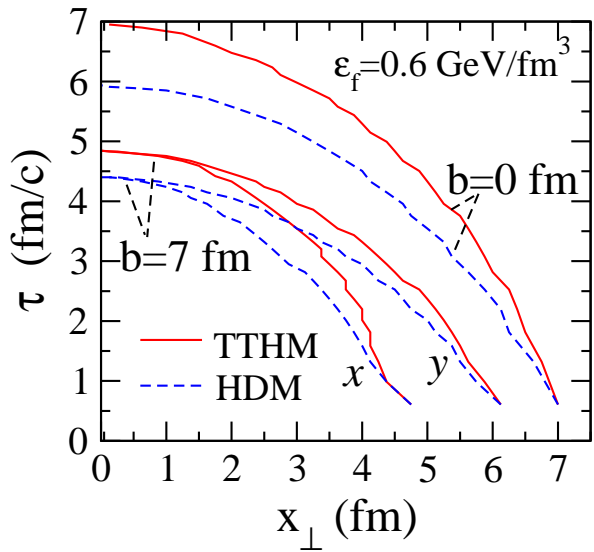,width=0.9\linewidth} 
\caption{The freeze-out time $\tau_{\rm f}(x,y)$ as a function of the
transverse coordinates $x$ and $y$, respectively, for a freeze-out 
energy density $\ve_\fo\EQ0.6$\,GeV/fm$^3$. The solid (dashed) 
lines are for TTHM (HDM), respectively. The upper set of curves is for
central collisions ($b\EQ0$\,fm) where the freeze-out surface is 
azimuthally symmetric. The lower set of curves corresponds to $b\EQ7$\,fm 
where we present cuts along the $x$ and $y$ axes, as indicated, to show 
the spatial azimuthal anisotropy at freeze-out. The impact parameter
$\bfat$ points in $x$ direction.} 
\label{F3}
\end{figure}

Before doing so, let us shortly comment on the rapidity densities $dN/dy$
and collision durations listed in Table~\ref{T1}. For identical initial 
conditions, the TTHM gives considerably lower multiplicity densities. 
This is mostly due to the higher freeze-out temperature in the TTHM which
implies that each gluon carries more energy (see Eq.~(\ref{epp})), so the
same freeze-out energy density $\ve_\fo$ translates into fewer gluons.
This argument overestimates the difference between the models, though; 
the actual difference is smaller since in the HDM the calculation of 
$dN/dy$ involves an integration over space-time rapidity $\eta$ which 
can be interpreted as taking an average over fireballs with different
flow rapidities $\eta{\,\ne\,}y$ and correspondingly reduced (redshifted) 
effective temperatures $T_{\rm eff}=T_\fo/\cosh(\eta{-}y)$. 

Longitudinal boost-invariance implies that the total transverse energy 
per unit rapidity, $dE_T/dy$, is independent of $y$. Due to the
absence of longitudinal pressure, no longitudinal work is done in the 
TTHM and $dE_T/dy$ is a constant of motion. This is different from 
the HDM where work done by the longitudinal pressure reduces $dE_T/dy$
with time. (Both statements were checked numerically.) \tref{T1} shows 
that, within numerical accuracy, the gluon multiplicity per unit 
rapidity, $dN/dy$, is also constant in time. While in ideal 
hydrodynamics (HDM) this is a simple consequence of entropy conservation 
(for boost-invariant systems $dS/dy$ is a constant of motion \cite{bj}, 
and for massless particles the entropy per particle is independent of 
temperature), there is no such simple reason for this observation in 
the TTHM which is far from local equilibrium. Detailed checks revealed 
that, within numerical accuracy, in the TTHM not only the transverse 
energy per particle, $E_T/N\EQ(dE_T/dy)/(dN/dy)$, but in fact the 
entire $\kt$-spectrum of the gluons is completely time-independent! 

If the system were transversally homogeneous and expanded only 
longitudinally, this would not be surprising: in the absence of
transverse gradients the conservation law \eref{t00} for $T^{00}$ 
simplifies to 
\be 
   \frac{\del (\t T^{00})}{\del \t} = 0 
\ee
with $T^{00}\EQ\ve\cosh^2\h$. In TTHM a constant product $\ve\t$ 
implies a constant temperature (see \eref{eps}) which, in the absence
of transverse flow, then leads to a time-independent transverse momentum
spectrum. In our case, however, the system expands in the transverse 
direction and cools. Our studies show that, in the longitudinal rest 
frame, the resulting loss in thermal transverse energy per particle 
is exactly compensated by the gain in transverse collective motion 
energy, in a way which exactly preserves the shape of the transverse 
momentum distribution. Although one should think that there must be 
a simple reason for this intriguing behaviour, we have not been able
to find a simple analytical proof and can thus only present the 
numerical evidence.

The longer freeze-out times in the TTHM simulations (see Table~\ref{T1})
are a reflection of the lack of longitudinal pressure. In the HDM the
pressure performs longitudinal work and thus causes a more rapid decrease
of the energy density with time than in the TTHM. This is further 
elucidated in Fig.~\ref{F3} where we show the freeze-out time 
$\t_\fo(x,y)$ as a function of position in the transverse plane. One
sees that, for identical initial conditions, the HDM simulations lead
to universally earlier freeze-out. This Figure also shows that for 
non-central Pb+Pb collisions ($b\EQ7$\,fm) the source at freeze-out is 
still larger in the $y$ direction perpendicular to the reaction plane
than in the reaction plane; the initial out-of-plane deformation of the
reaction zone thus has not yet fully disappeared when the dynamical 
evolution was stopped. This explains why the elliptic flow had not
yet saturated, see Fig.~\ref{F1} and Fig.~\ref{F9} below.

\subsection{Retuning the initial conditions for TTHM}
\label{sec5c}

\begin{figure}
\epsfig{file=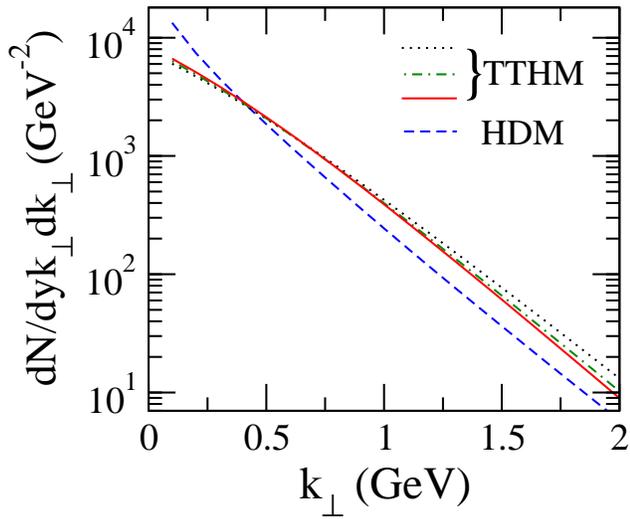,width=0.95\linewidth}
\caption{The gluon $\kt$ distribution from hydrodynamics (dashed line, 
with the parameters as given in the text) and TTHM (other lines).
The TTHM curves correspond to the following sets of parameters:
Dotted line: $\t_0\EQ6$\,fm/$c$, $\ve_0\EQ{e}_0/15$;
dot-dashed line: $\t_0\EQ7$\,fm/$c$, $\ve_0\EQ{e}_0/18$;
solid line: $\t_0\EQ7.8$\,fm/$c$, $\ve_0\EQ{e}_0/20$. Here 
$e_0\EQ23.0$\,GeV/fm$^3$ is the initial central energy density for $b\EQ0$
Au+Au collisions in the HDM.
} 
\label{F4} 
\end{figure}
 
Our TTHM model does not include the hadronization of gluons and hence 
does not allow us to compute hadron spectra; hence we cannot directly
compare the gluon multiplicity density $dN/dy$ and their $\kt$-distribution 
from TTHM at $\ve_\fo$ to any data. However, the relation of these 
quantities to the observable hadron spectra after hadronization is
not expected to depend on their dynamical history prior to reaching
the ``freeze-out'' point $\ve_\fo$. Thus, if we want to create TTHM 
solutions which are likely to lead, after hadronization, to hadron 
spectra with similar normalization and shape as the data, we can use 
the gluon rapidity densities and $\kt$-spectra from existing hydrodynamic 
calculations with initial conditions which {\em were} tuned to real data 
in \cite{khh}, evaluate them at the same value of $\ve_\fo$ where we stop 
the TTHM evolution, and retune the TTHM initial conditions such that they 
reproduce these HDM gluon spectra. With the TTHM spectra thus tied down 
to the final state in central ($b\EQ0$\,fm) collisions from the HDM,
we can then repeat our comparison of elliptic flow in the two models 
under more realistic boundary conditions.    

For the common decoupling point we chose in both TTHM and HDM the value
$\ve_\fo\EQ0.35$\,GeV/fm$^3$. This may appear somewhat low for still
using only gluon degrees of freedom (with an ideal gas EOS in the 
HDM case), but it has the advantage that then the hydrodynamic model with
the ideal massless EOS used here produces an elliptic flow $v_2$ of 
roughly the same magnitude as that obtained for pions in Ref.~\cite{khh} 
using a more realistic equation of state. Given the good agreement of 
those HDM calulations with the data \cite{khh}, we can thus pretend to 
be comparing directly to data when comparing the TTHM results to our 
HDM reference.

With this freeze-out energy density, our HDM reference gives the gluon 
$\kt$ (or $\mt$) spectrum shown in \fref{F4}, which integrates to a gluon
rapidity density at $y\EQ0$ of $dN/dy{\,\simeq\,}942$. Given the large
difference between the HDM and TTHM spectra for identical initial 
conditions shown in \fref{F2}, it is perhaps not surprising that we found
it rather difficult to reproduce the HDM spectrum within the TTHM: 
one needs to find a way to dramatically lower the radial flow (since
the freeze-out temperature is fixed by $\ve_\fo$ and is much higher in
TTHM than in HDM) and at the same time increase the normalization of
the spectrum in order to raise $dN/dy$.  

\begin{figure}
\epsfig{file=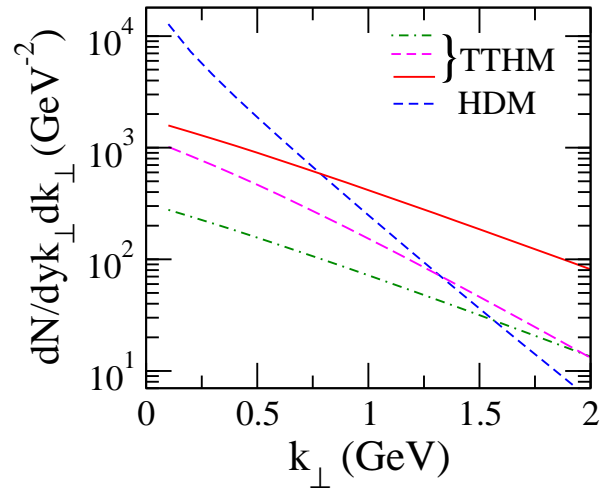,width=0.9\linewidth}
\caption{Variation of the gluon $\kt$ distribution with various TTHM 
model parameters. The solid and dashed lines repeat those from
\fref{F2} for $\ve_\fo$= 0.6 GeV/fm$^3$ for comparison. The dot-dashed 
line is obtained by starting the system earlier at $t_0\EQ0.1$\,fm/$c$.
The long-dashed line has a lower initial energy density 
$\ve_{\rm i}\EQ\ve_0/4$. In each case the remaining parameters are held 
fixed at the values used in \fref{F2}.}   
\label{F5} 
\end{figure}

There are essentially only three parameters in TTHM that we can play with: 
$\ve_0$, $\t_0$ and $\ve_\fo$. Examples of the effects of varying them are 
shown in \fref{F5}. The lines from \fref{F2} corresponding to 
$\ve_\fo\EQ0.6$\,GeV/fm$^3$ are repeated here for comparison. From 
\fref{F2} and \tref{T1} we learned already that changing $\ve_\fo$ 
has no influence on $dN/dy$ and little effect on $dN/dy\kt d\kt$. This
leaves us only with changes of the initial conditions $\ve_0$ and $\t_0$.
The dot-dashed curve in \fref{F5} shows that decreasing $\t_0$ from 0.6 to 
0.1\,fm/$c$ goes in the wrong direction, by reducing $dN/dy$ without 
having much effect on the slope of the $\kt$ spectrum. Reducing the 
initial energy density $\ve_0$ by a factor of four produces the 
long-dashed line in the figure. Due to the lower initial energy density
there is less time until freeze-out to produce radial flow, and the 
resulting spectrum is steeper, as desired; unfortunately, its normalization 
$dN/dy$ decreases, too. Combining the insights from these two trial runs,
we see that making the TTHM spectrum sufficiently steep requires still much
smaller initial energy densities $\ve_0$, combined with much larger
starting times $\tau_0$ for the TTHM transverse dynamical evolution, in 
order to increase $dN/dy$. This implies a long transverse thermalization 
time, which poses intrinsic consistency problems to be addressed later. 

Following this route, we find that the HDM gluon spectrum can be roughly 
reproduced with either one of the following three sets of parameters:
$\t_0\EQ6$\,fm/$c$ with $\ve_0\EQ{e}_0/15$, $\t_0\EQ7$\,fm/$c$ and 
$\ve_0\EQ{e}_0/18$, or $\t_0\EQ7.8$\,fm/$c$ and $\ve_0\EQ{e}_0/20$. The 
corresponding spectra are shown in \fref{F4}. These parameter sets are 
not unique and the agreement with the HDM spectra is not perfect, but 
what is clear is that a large reduction of $\ve_0$ by at least $1/15$ 
is essential in TTHM to obtain the much deeper HDM slope. 

\subsection{Elliptic flow from TTHM with retuned initial conditions}
\label{sec5d}

\begin{figure}
\epsfig{file=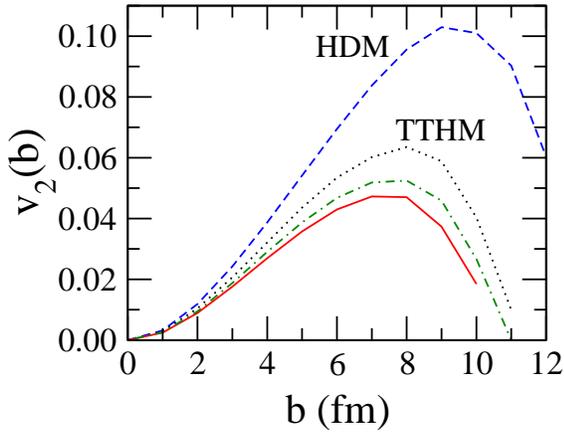,width=0.85\linewidth}
\caption{$v_2$ as a function of impact parameter from TTHM (solid) and 
HDM (dashed) simulations tuned to produce identical angle-averaged spectra 
$dN/dy\,\kt d\kt$. The parameters used and notations here are the same as 
those in \fref{F4}.} 
\label{F6} 
\end{figure}
 
Figure~\ref{F6} again shows the impact parameter dependence of $v_2$,
but now with retuned TTHM initial conditions as described in the previous
subsection. With the much lower $\ve_0$ forced upon us by the given slope 
of the final $\kt$ spectrum, there is much less time to generate transverse
flow. Figure~\ref{F6} shows that this leads not only to reduced radial
flow, as reflected in the steeper single-particle spectrum for central
collisions (see Fig.~\ref{F4}), but also cuts down on the elliptic flow
$v_2$ which now remains significantly below the HDM level. As the 
latter is representative of the data \cite{rs,hk}, we conclude that a 
model with only transversally thermalized momenta in the early collision 
stages cannot produce as much elliptic flow as required by experiment.

\begin{figure}
\epsfig{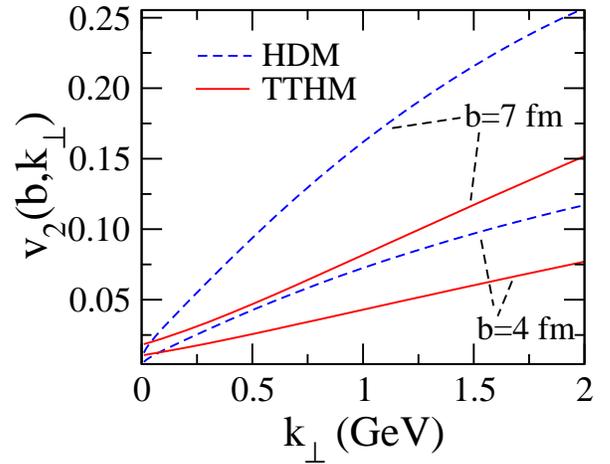}
\caption{$v_2$ as a function of $\kt$ for two impact parameters (as 
indicated) from TTHM (solid) and HDM (dashed) simulations giving 
identical angle-averaged spectra $dN/dy\,\kt d\kt$.} 
\label{F7} 
\end{figure}

\begin{figure}
\epsfig{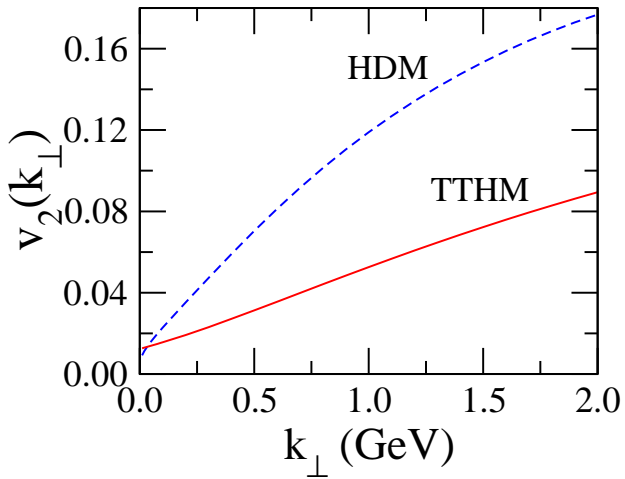}
\caption{Impact parameter averaged $v_2$ as a function of $\kt$ from 
the two models. The dashed line is from HDM, the solid line from TTHM 
with initial conditions tuned to produce the same angle-averaged 
transverse momentum spectrum at the same value of $\ve_f$.
} 
\label{F8} 
\end{figure}

In order to further strengthen this argument let us look at the $\kt$
dependence of $v_2$ and study (see Fig.~\ref{F7})
\be 
   v_2(\kt;b) = \frac{\int d\f\, \cos(2\f)\, \frac{dN(b)}{dy\,\kt d\kt d\f}} 
                     {\int d\f \,\frac{dN(b)}{dy\,\kt d\kt d\f} }  
\label{eq:v2bkt}
\ee
as well as its impact-parameter averaged (``minimum bias'' \cite{st1})
value (see Fig.~\ref{F8})
\be 
   v_2(\kt) = 
   \frac{\int b\, db\,\int d\f\, \cos(2\f)\, \frac{dN(b)}{dy\,\kt d\kt d\f}} 
        {\int b\, db\, \int d\f \,\frac{dN(b)}{dy\,\kt d\kt d\f}}  \;.  
\label{eq:v2kt}
\ee
We here select for our comparison with HDM one of the above three sets of 
initial parameters for TTHM, namely $\t_0\EQ7$\,fm/$c$ and 
$\ve_0\EQ{e}_0/18$, corresponding to the dot-dashed line in \fref{F4}.

Figures~\ref{F7} and \ref{F8} show that the $\kt$-slope of $v_2$ is much
lower than in the HDM. This is not only true for the impact parameter
averaged elliptic flow shown in Fig.~\ref{F8}, but holds universally
for all impact parameters; two examples are plotted in Fig.~\ref{F7}. 
For $v_2(\kt)$ from minimum bias events accurate data were published in 
\cite{st1,st2,rs} which follow essentially the dashed HDM line in 
Fig.~\ref{F8} and definitely exclude our TTHM. 

\begin{figure}
\epsfig{file=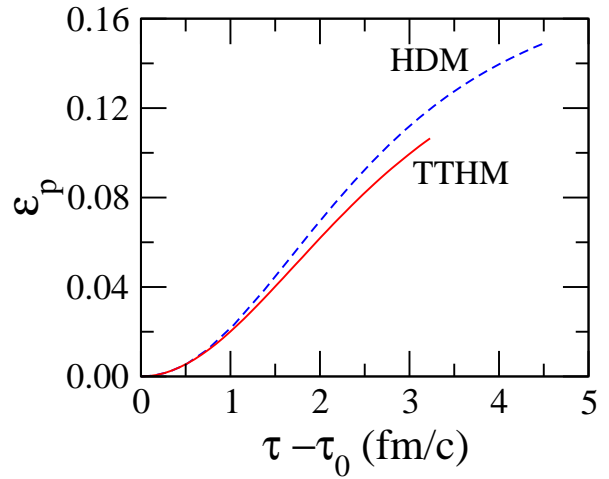,width=0.9\linewidth}
\caption{The asymmetry measure $\ve_p$ as a function of the duration of
the time evolution at $b=7$ fm with similar final particle spectra from
both models. Again the solid is from TTHM and the dashed line from
HDM.} 
\label{F9} 
\end{figure}

In Figure~\ref{F9} we show the time evolution of the momentum anisotropy 
\cite{ksh}
\be  
  \ve_p  = \frac{\lan T^{xx}-T^{yy} \ran}{\lan T^{xx}+T^{yy} \ran} 
\ee
where the angular brackets indicate an average over the transverse 
plane. At freeze-out, this momentum anisotropy is translated into 
elliptic flow $v_2$, where the coefficient between the two variables
depends on the particle rest mass \cite{ksh}. Figure~\ref{F9} shows 
that with the hard equation of state for non-interacting massless
gluons and RHIC-type initial conditions freeze-out actually happens 
before the momentum anisotropy saturates \cite{fn2} (and before the 
initial spatial anisotropy has fully disappeared). This is even more
true for the TTHM (with its even stiffer EOS of $\pt\EQ\frac{1}{2}\ve$) 
than for the HDM (where $P\EQ\frac{1}{3}\ve$): the TTHM freezes out
significantly earlier, even though about the same amount of 
transverse flow is generated.

\subsection{The low-$\kt$ limit of $v_2(\kt)$ for massless particles}
\label{sec5e}
 
\begin{figure}
\epsfig{file=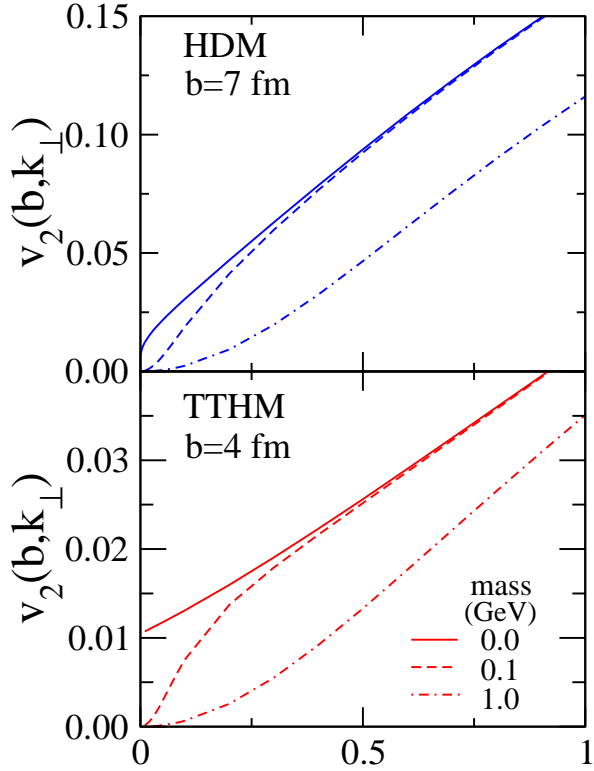,width=0.9\linewidth}
\caption{$v_2$ as a function of $\kt$ for massless and massive bosons
from the two models. The solid lines are for massless bosons while the
dashed and dot-dashed lines are for bosons with $m$=100\,MeV and $m$=1\,GeV,
respectively.} 
\label{F10} 
\end{figure}

A careful inspection of Figures~\ref{F7} and \ref{F8} shows that at small 
transverse momenta the shape of the $\kt$-differential elliptic flow 
$v_2(\kt)$ is different for TTHM and HDM. Figure~\ref{F10} further
shows that the small-$\kt$ behaviour of the elliptic flow coefficient
differs for massive and massless bosons. In Appendix~\ref{appb} we present
a detailed analytical treatment of the limiting behaviour of $v_2(\kt)$
for small transverse momenta. For massive particles, the elliptic flow
vanishes quadratically at $\kt\to0$,
\be
\label{mne0}
  v_2(\kt) = {\cal O}(\kt^2) \qquad \mbox{(massive particles)},
\ee
going over to an almost linear rise at higher $\kt$ (which eventually 
turns over to saturate at $v_2=1$ as $\kt\to\infty$ \cite{hkhrv}). The
transition from the quadratic rise at low $\kt$ to the quasi-linear 
behaviour at intermediate $\kt$ occurs around $\kt\sim m$. In detail 
the transition between these regimes is complex and depends on the ratio
$m/T$ of the rest mass to the freeze-out temperature (see Fig.~\ref{F10}):
for particles with $m/T<1$ the elliptic approaches the curve for massless
particles while the elliptic flow of heavier particles with $m/T\gg1$ 
remains always significantly below that of massless particles.

For massless bosons, the singularity of the Bose distribution at zero 
momentum leads to a qualitative change in the low-momentum limit of
the elliptic flow. For the TTHM, which lacks a thermal spread of the
longitudinal momenta, the elliptic flow of massless bosons (gluons
or photons) approaches a positive constant with finite positive slope:
\be
\label{v2-tthm}
  v_2(\kt) = a + b\kt + {\cal O}(\kt^2) 
  \quad \mbox{(TTHM, massless bosons)}.
\ee
In the HDM, the longitudinal thermal momentum smearing weakens the 
Bose singularity and causes $v_2$ to still vanish at zero transverse 
momentum, albeit just barely so: as $\kt\to0$, $v_2(\kt)$ vanishes 
with infinite slope:
\be
\label{v2-hdm}
  v_2(\kt) = \frac{const.}{\ln{\frac{T}{\kt}}}
  \qquad \mbox{(HDM, massless bosons)}.
\ee
It would be interesting to verify this prediction of the hydrodynamic
model for photons which are directly emitted from the expanding fireball.

\section{Conclusions}
\label{sec6}

We have tried to construct a macroscopic dynamical description of
the fireball expansion in relativistic heavy-ion collisions which
relaxes the usual assumption of complete local thermal equilibrium
which underlies the popular and successful hydrodynamical approach.
This addresses, on a macroscopic level, the observation that existing
parton-based microscopic kinetic models \cite{pcm,wong,molnar} seem 
unable to generate sufficiently fast thermalization to reach the 
hydrodynamic limit. We have tested a toy model where transverse
momenta are thermalized very quickly while the system is streaming 
freely in the longitudinal direction (i.e. without longitudinal momentum
transfer). The initial longitudinal motion was assumed to be 
boost-invariant, and the model preserves this property dynamically at 
later times.

Starting from the kinetic transport equation, we derived the corresponding
macroscopic equations of motion (TTHM) and solved them in parallel with 
the usual hydrodynamic (HDM) equations. Since our TTHM model is 
parton-based and lacks a description of hadronization, we decided to 
compare the two approaches just before the onset of hadronization. To make 
phenomenologically meaningful comparisons we used a HDM reference which 
had previously been shown to yield a good description of RHIC data when
allowed to evolve until hadron freeze-out.

The TTHM turns out to have an extremely stiff equation of state, 
$\pt\EQ\frac{1}{2}\ve$, which causes severe phenomenological problems:
in order to avoid the creation of too much radial transverse flow, which
would render the single-particle spectra much flatter than observed, 
while preserving the total multiplicity, we had to start the TTHM evolution
very late, at already quite low energy densities. This implies a very long
transverse thermalization time scale (and, of course, an even longer
longitudinal one). While this cannot a priori be excluded theoretically,
it causes consistency problems with our implementation which assumes
that transverse collective dynamics sets in only after we start the TTHM
evolution (i.e. only after thermalization). This is not realistic: even 
before complete thermalization of the transverse momenta the system
will start to evolve in the transverse directions, since nothing confines
it. This pre-equilibrium transverse dynamics will likely have a collective 
component affecting the transverse momentum spectra, necessitating an even
further shift of the starting point for TTHM evolution. 

We have not followed this issue any further, since there is also a severe 
problem when comparing the TTHM predictions with the elliptic flow data,
which in our opinion eliminates it as a viable model. For the same radial 
flow, constrained by the slope of the single-particle spectra, TTHM 
provides only about half as much elliptic flow as the hydrodynamical model 
and thus significantly less than that required by the RHIC data 
\cite{st1,st2,rs,lacey}. This is the central result from the present 
study, and it strengthens the previously made argument \cite{khh,hk}
that the large elliptic flow measured at RHIC can only be reproduced
by models which assume very rapid thermalization of the momentum spectra
{\em in all three dimensions}. Partial thermalization, such as the
effectively only two-dimensional thermalization in the TTHM model, is 
not enough to generate the observed elliptic flow. This leaves us with
the continued challenge of identifying the microscopic mechanisms within 
QCD which can cause such rapid three-dimensional thermalization.

\section*{Acknowledgements}

S.W. thanks Peter Kolb for the FORTRAN code for calculating the nuclear 
overlap function which was used to initialize the TTHM time evolution. 
This work was supported by the U.S. Department of Energy under Contract No. 
DE-FG02-01ER41190.

\bigskip

\appendix
\section{Decomposing the Energy-Momentum Tensor} 
\label{appa} 

To calculate the coefficients in the decomposition (\ref{eq:tmn-ten})
of the energy-momentum tensor we find it easiest to use the form
(\ref{eq:dist}) for the distribution function in \eref{eq:tmn-def}. 
Assuming massless particles (gluons), we write the four-vector $k^\m$ as 
\be  
   k^\m = \kt (\cosh y, \cos \f, \sin \f, \sinh y)\,.
\ee
We integrate over $d^3k/k_0{\EQ}dy\,\kt d\kt\,d\phi$ in \eref{eq:tmn-def} 
by first doing the trivial $\h$-integration and then integrating over 
$\kt$. After integration over $\h$ the exponent of the Bose factor in 
\eref{eq:dist} reduces to 
\be 
   \frac{k{\cdot}u}{T} = \frac{\gt\kt}{T}(1-\hktv{\cdot}\vtv)
   = \frac{\gt\kt}{T} (1-\vt\cos(\f{-}\f_s))
\ee
where $\f{-}\f_s$ is the angle between $\ktv$ (azimuthal angle $\f$) and 
$\vtv$ (azimuthal angle $\f_s$). The $\kt$-integration is then easily 
done using
\be\label{A1}  
     \int^\infty_0 \frac{\kt^n d\kt}{e^{\a \kt}-1} = 
     \frac{n!\,\z(n+1)}{\a^{n+1}}\,. 
\ee
In our case $\alpha\EQ\gt(1-\vt\cos(\f{-}\f_s))/T$ and $n\EQ3$, and the 
relevant zeta function is $\z(4)= \p^4/90$. In this way we arrive at
\be 
    T^{\m\n} = \frac{\t_0}{\t} \frac{\n}{\gt^5} \frac{\p^2 T^4}{60}
    \int_{-\pi}^\pi \frac{d\f}{2\p} \frac{M^{\m\n}}{(1-\vt \cos(\f{-}\f_s))^4} 
\label{eq:tmn} 
\ee
with
\onecolumngrid
\be 
    M^{\m\n} = 
    \begin{pmatrix}  
    \cosh^2 \h        &\quad\cos\f\cosh\h   &\quad \sin\f\cosh \h
                      &\quad \cosh\h\sinh\h 
\\
    \cos \f \cosh \h  &\quad \cos^2\f        &\quad \cos\f\sin\f  
                      &\quad \cos\f\sinh\h  
\\
    \sin  \f \cosh \h &\quad \cos\f\sin\f    &\quad \sin^2\f        
                      &\quad \sin\f\sinh\h  
\\ 
    \cosh \h \sinh \h &\quad \cos\f\sinh\h   &\quad \sin\f\sinh\h 
                      &\quad \sinh^2\h 
    \end{pmatrix}  \;.
\ee
The remaining azimuthal integration is done by shifting the
integration variable to $\theta\EQ\f-\f_s$, using
\be  
     \cos \f = \cos\theta\cos\f_s - \sin\theta\sin\f_s\,,\qquad
     \sin \f = \sin\theta\cos\f_s + \cos\theta\sin\f_s    
\label{eq:trig-id} 
\ee
in the numerator $M^{\m\n}$, and exploiting the formula
\be 
  \int^{\p}_{-\p} \frac{d\theta}{2\p} \frac{1}{(1{-}\vt\cos\theta)^n} 
     = \frac{1}{(1{-}\vt^2)^{n/2}} 
         P_{n{-}1}\Bigl(\frac{1}{\sqrt{1{-}\vt^2}}\Bigr) 
     = \gt^n\,P_{n{-}1}(\gt)\,,
\label{eq:phi-int}
\ee
where the $P_n$ are Legendre polynomials. We need
\be  
   \gt^2 P_1(\gt) = \gt^3\,,\qquad
   \gt^3 P_2(\gt) = \gt^5\left(1+\mbox{$\frac{1}{2}$} \vt^2\right)\,,\qquad
   \gt^4 P_3(\gt) = \gt^7\left(1+\mbox{$\frac{3}{2}$} \vt^2\right)\,.
\ee
$M^{00}$, $M^{03}{\EQ}M^{30}$, and $M^{33}$ have no $\f$-dependence
and the corresponding angular integrals thus give
\be  
   \int^{\p}_{-\p} \frac{d\f}{2\p} \frac{1}{(1-\vt \cos(\f-\f_s))^4} 
   = \int^{\p}_{-\p} \frac{d\theta}{2\p} \frac{1}{(1-\vt \cos\theta)^4} 
   = \gt^4 P_3(\gt) 
   = \gt^7 \Big (1+\mbox{$\frac{3}{2}$} \vt^2 \Big ) \,.
\ee
For $M^{01}{\EQ}M^{10}$ and $M^{13}{\EQ}M^{31}$ we need
\bea 
     \int^{\p}_{-\p} \frac{d\f}{2\p} \frac{\cos \f}{(1-\vt\cos(\f{-}\f_s))^4}
 &=& \int^{\p}_{-\p} \frac{d\theta}{2\p} 
     \frac{\cos\theta\cos\f_s -\sin\theta\sin\f_s}{(1-\vt \cos\theta)^4} 
   = \frac{\cos\f_s}{\vt} \int^{\p}_{-\p} \frac{d\theta}{2\p}  
     \frac{1-(1{-}\vt\cos\theta)}{(1-\vt\cos\theta)^4}
\nonum\\ 
 &=& \frac{v_x}{\vt^2} \Big (\gt^4 P_3(\gt) - \gt^3 P_2(\gt) \Big )
     = \mbox{$\frac{1}{2}$} \gt^7 v_x (4+\vt^2) \; . 
\label{eq:c-phi-int}
\eea
In the last step on the first line we dropped the vanishing term 
$\sim\sin\theta$ and also added and subtracted a term $1/\vt$ in 
the numerator. In the second line we used $\cos\f_s{\EQ}v_x/\vt$. For 
the remaining components we proceed similarly. The energy-momentum 
tensor thus takes the form 
\be 
   T^{\m\n} = \frac{\t_0}{\t} \n_g \frac{\p^2 T^4}{60} \gt^2 
    \left ( \begin{matrix}
    (1+\frac{3}{2} \vt^2) \cosh^2 \h 
  & \half v_x (4+\vt^2) \cosh \h          & \half v_y (4+\vt^2) \cosh \h    
  & (1+\frac{3}{2} \vt^2) \cosh \h \sinh \h                           \\
    \half v_x (4+\vt^2) \cosh \h          & \half (1+4 v_x^2-v_y^2) 
  & \frac{5}{2} v_x v_y                   & \half v_x (4+\vt^2) \sinh \h   \\
    \half v_y (4+\vt^2) \cosh \h          & \frac{5}{2} v_x v_y       
  & \half (1-v_x^2+4 v_y^2)               & \half v_y (4+\vt^2) \sinh \h   \\
    (1+\frac{3}{2} \vt^2) \cosh \h \sinh \h
  & \half v_x (4+\vt^2) \cosh \h          & \half v_y (4+\vt^2) \sinh \h
  & (1+\frac{3}{2} \vt^2) \sinh^2 \h 
    \end{matrix} \right )   \;. 
\nonum
\ee
Contracting this result once with each of the vectors $u^\m,\,n^\m,$ and 
$m^\m$ we obtain
\be
  T^{\m\n} u_\n = \frac{\t_0}{\t}\n_g \frac{\p^2 T^4}{60} \gt^3 
    \left (\begin{smallmatrix}
           (1-\half \vt^2 -\half \vt^4 ) \cosh \h    \\
           \frac{3}{2} v_x (1-\vt^2)                 \\ 
           \frac{3}{2} v_y (1-\vt^2)                 \\ 
           (1-\half \vt^2 -\half \vt^4 ) \sinh \h    
           \end{smallmatrix} \right ), \quad   
  T^{\m\n} m_\n = - \frac{\t_0}{\t} \n_g\frac{\p^2 T^4}{60} \gt^3 
    \left (\begin{smallmatrix}
           (1-\vt^2)\vt  \cosh \h                  \\
           v_x (1-\vt^4)/2\vt            \\ 
           v_y (1-\vt^4)/2\vt            \\ 
           (1-\vt^2)\vt  \sinh \h      
           \end{smallmatrix} \right ),\quad
  T^{\m\n} n_\n = 0.   
\nonum
\ee
A second contraction then yields the coefficients $A$ through $H$ in
\eref{eq:tmn-ten} as given in equations (\ref{eps})-(\ref{nTm}).

\vspace*{0.4cm}

\twocolumngrid

\section{The small transverse momentum limit of elliptic flow} 
\label{appb} 

In this Appendix we derive the small-$k_\perp$ limit of the elliptic
flow coefficient $v_2$ for TTHM and HDM, for massive and massless 
particles. Suppressing the impact parameter dependence which is
irrelevant here, we write \eref{eq:v2bkt} in the form
\be 
   v_2(\kt) = \frac{\cal{N}}{\cal{D}}\,,
\label{B1}
\ee
where the numerator ${\cal{N}}$ and denominator ${\cal{D}}$ are obtained
from the Cooper-Frye formula (\ref{eq:cf}). For massive particles we must
generalize \eref{fo-norm} by writing the first term as $\mt\cosh(y{-}\eta)$ 
and use 
\be
\label{B2}
   k\cdot u = \gt \bigl(\mt\cosh(y{-}\eta) - \ktv{\cdot}\vtv\bigr).
\ee
For the TTHM the distribution function is given by \eref{eq:dist}, and 
we obtain for particles at midrapidity $y\EQ0$
\be
\label{B3}
  \left\{{\cal N}\atop{\cal D}\right\} =
  \frac{\n_g\t_0}{(2\pi)^2}
  \int\frac{d^2\xt}{\gt}\int_{-\infty}^\infty d\eta\,\delta(\eta)
  \left\{{\cal N}_\phi\atop{\cal D}_\phi\right\} \,, 
\ee
where
\be
\label{B4}
  \left\{{\cal N}_\phi\atop{\cal D}_\phi\right\} =
  \int_{-\pi}^\pi\frac{d\phi}{2\pi}
  \left\{\cos(2\phi)\atop 1\right\}
  \frac{\mt\cosh\eta{-}\ktv{\cdot}\nabla_\perp\t_\fo}
       {e^{\gt(\mt\cosh\eta{-}\ktv{\cdot}\vtv)/T}{-}1}.
\ee
Here $\gt$ and $\t_\fo$ are functions of the position in the transverse 
plane, $\gt\EQ\gt(\xtv)$ and $\t_\fo\EQ\t_\fo(\xtv)$. For the HDM we 
instead use \eref{equil} and get
\be
\label{B5}
  \left\{{\cal N}\atop{\cal D}\right\} =
  \frac{\n_g}{(2\pi)^2}
  \int d^2\xt \,\t_\fo \int_{-\infty}^\infty d\eta
  \left\{{\cal N}_\phi\atop{\cal D}_\phi\right\}\,.  
\ee
The main difference (except for the different weighting of the 
integration over the transverse plane $d^2\xt=r\,dr\,d\f_s$) is 
the nontrivial $\eta$-integration in the HDM case.

\medskip
\noindent
{\bf 1.} {\em Massive particles:} For non-zero rest mass, $m\ne0$, we 
can expand in both cases the $\phi$-integrand for small $\kt$. Keeping 
only terms up to first order in $\kt$ we find
\bea
\label{B6}
  &&\left\{{\cal N}_\phi\atop{\cal D}_\phi\right\} \approx
  \frac{m\cosh\eta}{e^{\gt m\cosh\eta/T}-1}
  \int_{-\pi}^\pi\frac{d\phi}{2\pi}
  \left\{\cos(2\phi)\atop 1\right\}
\\
  &&\times
  \left(1-\frac{\ktv{\cdot}\nabla_\perp\t_\fo}{m\cosh\eta}\right)
  \left(1+\frac{\ktv{\cdot}\vtv}{T}
          \frac{\gt}{1-e^{-\gt m\cosh\eta/T}}\right).
\nonumber
\eea
Higher orders of $\kt$ come with higher inverse powers of $T$ or
$m\cosh\eta$; the corresponding $\eta$-integrals thus are all finite
(trivially so in the TTHM), and the $\kt$-expansion is well-behaved. 

As in Appendix A we write $\ktv{\cdot}\vtv=\kt\vt\cos(\f{-}\f_s)$
 $\equiv \kt\vt\cos\theta$ and use $-\p\leq\theta\leq\p$ as integration
variable. Then, using \eref{eq:trig-id},
\be
\label{B7}
  \hktv\cdot\nabla_\perp\t_\fo = (\partial_r\t_\fo) \cos\theta +
  (\partial_\f\t_\fo)\sin\theta,
\ee
where $\partial_r\t_\fo\EQ\hvtv{\cdot}\nabla_\perp\t_\fo$ and 
$\partial_\f\t_\fo\EQ\hzv{\cdot}(\hvtv{\times}\nabla_\perp\t_\fo)$.
Here $\hzv$ is the unit vector along the beam direction, and 
$\hvtv$ coincides with the radial unit vector, 
$\hvtv\EQ(\cos\f_s,\sin\f_s)$. We also need 
\bea
\label{B8}
 \cos(2\f) &=& \cos(2\f_s)(\cos^2\theta-\sin^2\theta) 
\nonumber\\
         &-& 2 \sin(2\f_s)\sin\theta\cos\theta.
\eea
After these manipulations the angular integrations in \eref{B6} are easily
performed. Due to symmetric integration over the full circle, only terms 
containing even powers of $\cos\theta$ and/or $\sin\theta$ survive. We find
\be
\label{B9}
 {\cal N}_\phi = {\cal O}(\kt^2),\qquad 
 {\cal D}_\phi = a + b\kt +{\cal O}(\kt^2),
\ee
where $a$ and $b$ are integrable functions of $\xtv$ and $\eta$. As a 
result, {\em for massive particles} $v_2(\kt)\EQ{\cal O}(\kt^2)$, i.e. the
elliptic flow coefficient vanishes with zero slope when $\kt{\,\to\,}0$, as
first observed by Danielewicz \cite{pd}.

\medskip
\noindent
{\bf 2.} {\em Massless bosons in the TTHM:} For massless bosons at 
$y\EQ0$ the angular integrals ${\cal N}_\phi$, ${\cal D}_\phi$ in 
\eref{B4} reduce to
\bea
\label{B10}
  \left\{{\cal N}_\phi\atop{\cal D}_\phi\right\} &=&
  \frac{T}{\gt}\int_{-\pi}^\pi\frac{d\phi}{2\pi}
  \left\{\cos(2\phi)\atop 1\right\}
  \frac{1-\frac{\hktv{\cdot}\nabla_\perp\t_\fo}{\cosh\eta}}
       {1-\frac{\hktv{\cdot}\vtv}{\cosh\eta}}
\nonumber\\
  &&\times
  \frac{\frac{\gt\kt}{T}(\cosh\eta-\hktv{\cdot}\vtv)}
       {e^{\frac{\gt\kt}{T}(\cosh\eta-\hktv{\cdot}\vtv)}-1}.
\eea
We can try to expand the last factor for small $\kt$ by using
\be
\label{B11}
 \frac{x}{e^x-1}=1 -\frac{x}{2}+{\cal O}(x^2).
\ee
However, since in this expression the expansion parameter is 
$x=\frac{\gt\kt}{T}(\cosh\eta{-}\hktv{\cdot}\vtv)$, each additional 
power of $\kt$ brings in another factor $\cosh\eta$. This is no problem for
the TTHM with its factor $\delta(\eta)$ under the $\eta$-integral, but
for the HDM the $\eta$-integrals of the expansion coefficients diverge,
rendering the expansion meaningless. For the HDM thus the $\eta$-integration
must be performed {\em before} we expand for small $\kt$ (see {\bf 3.} 
below). 

For the TTHM, on the other hand, we can continue along this direction. 
Using the same manipulations as in {\bf 1.} above, we find up to linear 
order in $\kt$
\bea
\label{B12}
  &&\left\{{\cal N}_\phi\atop{\cal D}_\phi\right\} \approx
\nonumber\\
  &&\quad
  \frac{T}{\gt}\int_{-\pi}^\pi\frac{d\theta}{2\pi}
  \left\{\cos(2\f_s)\cos(2\theta)-\sin(2\f_s)\sin(2\theta)\atop 1\right\}
\nonumber\\
  &&\qquad\times
  \frac{1-(\partial_r\t_\fo) \cos\theta - (\partial_\f\t_\fo)\sin\theta}
       {1-\vt\cos\theta}
\nonumber\\
  &&\qquad\times
  \left(1-\frac{\gt\kt}{2T}(1{-}\vt\cos\theta)\right).
\eea
We use \eref{eq:phi-int} to derive the following table of integrals:
\bea
\label{B13}
  && \int_{-\pi}^\pi\frac{d\theta}{2\pi}\frac{1}{1{-}\vt\cos\theta} 
     =\gt,
\nonumber\\
  && \int_{-\pi}^\pi\frac{d\theta}{2\pi}\frac{\cos\theta}
                                             {1{-}\vt\cos\theta} 
     =\frac{\gt{-}1}{\vt},
\nonumber\\
  && \int_{-\pi}^\pi\frac{d\theta}{2\pi}\frac{\cos^2\theta}
                                             {1{-}\vt\cos\theta} 
     =\frac{\gt{-}1}{\vt^2},
\nonumber\\
  && \int_{-\pi}^\pi\frac{d\theta}{2\pi}\frac{\sin^2\theta}
                                             {1{-}\vt\cos\theta} 
     =\frac{\gt{-}1}{\gt\vt^2},
\nonumber\\
  && \int_{-\pi}^\pi\frac{d\theta}{2\pi}\frac{\cos(2\theta)}
                                             {1{-}\vt\cos\theta} 
     =\frac{1}{\gt}\left(\frac{\gt{-}1}{\vt}\right)^2,
\nonumber\\
  && \int_{-\pi}^\pi\frac{d\theta}{2\pi}\frac{\sin^2\theta\cos\theta}
                                             {1{-}\vt\cos\theta} 
     =\frac{1}{2\vt}\left(\frac{\gt{-}1}{\gt\vt}\right)^2,
\nonumber\\
  && \int_{-\pi}^\pi\frac{d\theta}{2\pi}\frac{\cos(2\theta)\cos\theta}
                                             {1{-}\vt\cos\theta} 
     =\frac{1}{\gt\vt}\left(\frac{\gt{-}1}{\vt}\right)^2,
\eea
and find
\bea
\label{B14}
  {\cal N}_\f&=&T\left(\frac{\gt{-}1}{\gt\vt}\right)^2
  \biggl[\,\cos(2\f_s)\left(1-\frac{\partial_r\t_\fo}{\vt}\right)
\nonumber\\
  &&\qquad\qquad\quad\,\
      +\sin(2\f_s) \frac{\partial_\f\t_\fo}{\gt\vt}\biggr] 
      +{\cal O}(\kt^2),
\nonumber\\
  {\cal D}_\f&=&
   T\left(1-\frac{\gt{-}1}{\gt\vt} (\partial_r\t_\fo)\right)
  -\frac{\kt}{2}+{\cal O}(\kt^2).
\eea
Now both the numerator and denominator approach nonzero constants as
$\kt\to0$. Since the fireball center freezes out later than its edge, the 
term $\sim(\partial_r\t_\fo)$, after integration over the transverse plane,
contributes with a negative sign, and the $\kt$-independent first terms
in ${\cal N}_\f$ and ${\cal D}_\f$ are correspondingly positive. As a 
result, the small-$\kt$ expansion of the elliptic flow coefficient takes 
the form 
\bea
\label{B15}
  v_2(\kt)&=&\frac{{\cal N}}{{\cal D}}=\frac{a}{1-b\kt} +{\cal O}(\kt^2)
\nonumber\\
  &=& a + ab\kt + {\cal O}(\kt^2),
\eea
where $a$ and $b$ are positive constants arising from the integration over 
the transverse profiles of the freeze-out time and flow velocity. Thus, as 
$\kt\to 0$, $v_2$ approaches a positive constant value with nonzero positive 
slope, as confirmed by Figs.~\ref{F7} to \ref{F9}.

\medskip
\noindent
{\bf 3.} {\em Massless bosons from the HDM:} In the last step we discuss 
the corresponding limit for massless bosons from the HDM, in order to 
understand the surprising behaviour shown in the upper part of 
Fig.~\ref{F9}. As already mentioned, in this case the $\eta$-integration 
must be performed {\em before} expanding in $\kt$. Doing this integration 
first, we rewrite \eref{B5} as
\bea
\label{B16}
  &&\left\{{\cal N}\atop{\cal D}\right\} =
  \frac{\n_g}{(2\pi)^2}
  \int r\,dr\,d\f_s \,\frac{\t_\fo(r,\f_s)}{\gt(r,\f_s)}
\nonumber\\
  &&\quad\times
  \int_{-\pi}^\pi\frac{d\theta}{2\pi}
  \left\{\cos(2\f_s)\cos(2\theta)-\sin(2\f_s)\sin(2\theta)\atop 1\right\}
\nonumber\\
  &&\quad\times 
  \int_{-\infty}^\infty d\eta
  \frac{\xi\left(\cosh\eta-\hktv{\cdot}\nabla_\perp\t_\fo\right)}
       {e^{\xi\left(\cosh\eta-\hktv{\cdot}\vt\right)}
        -1},
\eea
where all the dependence on the magnitude of $\kt$ is hidden in the variable
\be
\label{B17}
 \xi = \frac{\gt\kt}{T}.
\ee
Let us denote the result of the $\eta$-integral by $I_\eta$. Expanding
the Bose-Einstein distribution in a power series and performing the
$\eta$-integration term by term we find
\bea
\label{B18}
  I_\eta &=& \xi\sum_{n=1}^\infty e^{n\vt\xi\cos\theta}
             \int_{-\infty}^\infty d\eta\,e^{-n\xi\cosh\eta}
\nonumber\\
  &&\qquad\qquad\times 
    \bigl(\cosh\eta - (\partial_r\t_\fo) \cos\theta 
                           - (\partial_\f\t_\fo)\sin\theta\bigr)
\nonumber\\
  &=& 2\, \xi\sum_{n=1}^\infty e^{n\vt\xi\cos\theta} \Bigl[{\rm K}_1(n\xi)
\nonumber\\
  &&\qquad 
  - \bigl(\cos\theta(\partial_r\t_\fo)+\sin\theta(\partial_\f\t_\fo)\bigr)
  {\rm K}_0(n\xi)\Bigr],
\eea
where the K$_\n$ are modified Bessel functions of the second kind. The 
integration over the momentum-space angle $\theta$ can now be performed, 
too, yielding modified Bessel functions of the first kind I$_\n$:
\be
\label{B19}
  \int_{-\pi}^\pi\frac{d\theta}{2\pi}\,\cos(\n\theta)\,e^{z\cos\theta} 
     = {\rm I}_\n(z),
\ee
where $z\EQ{n}\vt\xi$. After some simple manipulations using 
trigonometric identities we find for the elliptic flow coefficient
\be
\label{B20}
  v_2(\kt) = \frac{\sum_{n=1}^\infty \int r\,dr\,d\f_s\,
             \frac{\t_\fo(r,\f_s)}{\gt(r,\f_s)}\, {\cal N}_n(r,\f_s;\kt)}
                  {\sum_{n=1}^\infty \int r\,dr\,d\f_s\,
             \frac{\t_\fo(r,\f_s)}{\gt(r,\f_s)}\, {\cal D}_n(r,\f_s;\kt)},
\ee
where
\bea
\label{B21}
  {\cal N}_n &=& \cos(2\f_s)\Bigl[\xi\,{\rm I}_2(n\vt\xi)\,{\rm K}_1(n\xi)
\nonumber\\
  &&\qquad\quad
  -\frac{\partial_r\t_\fo}{2}\,\xi
   \Bigl({\rm I}_1(n\vt\xi)+{\rm I}_3(n\vt\xi)\Bigr){\rm K}_0(n\xi)\Bigr]
\nonumber\\
  &+&\sin(2\f_s) \frac{\partial_\f\t_\fo}{2}\,\xi 
     \Bigl({\rm I}_1(n\vt\xi)-{\rm I}_3(n\vt\xi)\Bigr){\rm K}_0(n\xi)\,,
\nonumber\\
  {\cal D}_n &=& \xi\,{\rm I}_0(n\vt\xi)\,{\rm K}_1(n\xi)
  -(\partial_r\t_\fo)\, \xi\, {\rm I}_1(n\vt\xi)\,{\rm K}_0(n\xi)\,.
\nonumber\\
\eea
These expressions generalize the result given in \cite{hkhrv} to the 
general case of a freeze-out time $\t_\fo$ which depends on the 
transverse position $\xtv$. 

We can now study the small-momentum limit of $v_2$ by letting $\xi\to0$.
Let us first consider the Boltzmann approximation which corresponds to 
keeping only the terms ${\cal N}_1$ and ${\cal D}_1$ in \eref{B20}. Using
the small-$\xi$ expansions of the Bessel functions one finds that in
this approximation 
\be
\label{22}
  \lim_{\kt\to0}v_2^{\rm Boltz}(\kt)\sim 
  \frac{\kt^2}{T^2}\ln\frac{\kt}{T}.
\ee
For massless Boltzmann particles, the elliptic flow from the HDM thus
goes to zero with zero slope, but by a factor $\ln(\kt/T)$ more slowly 
than for massive particles, see point {\bf 1.} above.

For bosons we must sum over all $n$ in \eref{B20}. For $\kt\to0$, i.e. 
$\xi\to0$, the sum over $n$ becomes the discrete representation of
an integral: defining 
\be
\label{B23}
  f_{\vt}(\zeta)\equiv{\rm I}_\n(\vt\zeta)\,{\rm K}_\mu(\zeta),
\ee
we have
\bea 
\label{B24}
  &&\xi\sum_{n=1}^\infty {\rm I}_\n(n\vt\zeta)\,{\rm K}_\mu(n\zeta) =
  \xi\sum_{n=1}^\infty f_{\vt}(n\zeta) 
\nonumber\\
  &&\qquad\quad\longrightarrow
  \int_{\xi/2}^\infty f_{\vt}(\zeta)\,d\zeta.
\eea
By expanding the Bessel functions for large arguments one sees that
for large $\zeta$
\be
\label{B25}
  f_{\vt}(\zeta)\approx \frac{1}{2\sqrt{\vt}\zeta}\,e^{-(1-\vt)\zeta},
\ee
such that the integral in \eref{B24} clearly converges at the upper end.
The $\kt$-dependence of $v_2$ can now be extracted by studying the
dependence of this integral on its lower limit $\xi/2=\gt\kt/2T$. To do
so we employ the following trick: We split the integral at an arbitrary
intermediate scale $\lambda$,
\bea
\label{B26}
  \int_{\xi/2}^\infty f_{\vt}(\zeta)\,d\zeta &=&
  \int_{\xi/2}^\lambda f_{\vt}(\zeta)\,d\zeta +
  \int_\lambda^\infty f_{\vt}(\zeta)\,d\zeta
\nonumber\\
  &=& \int_{\xi/2}^\lambda f_{\vt}(\zeta)\,d\zeta +
  F(\lambda;\vt),
\eea
where $F(\lambda;\vt)$ is a finite number whose dependence on $\lambda$
cancels against the $\lambda$-dependence of the first integral, 
irrespective of how we choose $\lambda$. By taking $\xi/2<\lambda\ll 1$
we can evaluate the first integral analytically by making use of the
expansion of the Bessel functions for small arguments, keeping only
the leading terms. In the limit $\xi\to0$ we find in this way
\bea
\label{B27}
  && \int_{\xi/2}^\lambda {\rm I}_2(\vt\zeta)\,{\rm K}_1(\zeta)\,d\zeta
     \longrightarrow \frac{\lambda^2\vt^2}{16}\,,
\nonumber
\eea
\bea
  && \int_{\xi/2}^\lambda {\rm I}_1(\vt\zeta)\,{\rm K}_0(\zeta)\,d\zeta
     \longrightarrow \frac{\vt}{16}\xi^2\ln(a\xi)+C(\lambda)\,,
\nonumber\\
  && \int_{\xi/2}^\lambda {\rm I}_3(\vt\zeta)\,{\rm K}_0(\zeta)\,d\zeta
     \longrightarrow 0\,,
\nonumber\\
  && \int_{\xi/2}^\lambda {\rm I}_0(\vt\zeta)\,{\rm K}_1(\zeta)\,d\zeta
     \longrightarrow \ln\frac{2\lambda}{\xi}+C'(\lambda)\,,
\eea
where $a$ is a $\lambda$-independent constant while $C,\,C'$ depend on
$\lambda$ but combine with the corresponding constants $F(\lambda;\vt)$ 
to finite, $\lambda$-independent constants. Inserting these results into
Eqs.~(\ref{B20}) and (\ref{B21}) we finally obtain
\be
\label{B28}
  \lim_{\kt/T\to0} v_2(\kt) = \frac{const.}{\ln{\frac{T}{\kt}}}\,,
\ee
which approaches zero with infinite slope. This is exactly the behaviour
seen in the upper panel of Fig.~\ref{F9} for massless bosons.



\begin{thebibliography}{99}
\bibitem{fn1}
This assumption is not unreasonable because flow ani\-so\-tro\-pies require
anisotropic pressure gradients for building up, but at the same time 
they reduce the latter dynamically \cite{s97}. The necessary spatial 
transverse anisotropies thus exist only during the very early collision 
stages, and we here assume that the time needed to eliminate them is
shorter than the longitudinal thermalization time.
\bibitem{fn2}
  As shown in Fig.~1 of \cite{ksrh}, the equation of state of an ideal
  massless gluon gas is so stiff that at RHIC energies the fireball 
  freezes out dynamically (i.e. driven by strong transverse flow) before 
  the spatial deformation of the reaction zone in noncentral collisions 
  has fully disappeared. Therefore, with such an equation of state, 
  elliptic flow does not fully saturate before decoupling. Nonetheless,
  the same Figure shows \cite{ksrh} that more than half of the maximally 
  possible elliptic flow develops before the decoupling energy density is 
  reached.
\end{thebibliography}
\end{document}